\documentclass[a4paper,11pt,twoside]{article}
\usepackage[a4paper,left=2.2cm,right=2.2cm,top=2.2cm,bottom=2.2cm]{geometry}
\usepackage[english]{babel}
\usepackage{setspace}
\usepackage{mathrsfs,hyperref, algorithm, algorithmic}
\usepackage{url,breakurl}
\usepackage[]{amsmath}
\usepackage{amsthm}
\usepackage{fix-cm}
\usepackage[]{amssymb}
\usepackage[]{latexsym}
\usepackage[latin1]{inputenc}
\usepackage[right]{eurosym}
\usepackage[T1]{fontenc}
\usepackage[]{graphicx}
\usepackage{wrapfig}
\usepackage[]{epsfig}
\usepackage{fancyhdr}
\usepackage{bbm}
\usepackage{pstricks}
\usepackage{multirow}
\usepackage[font=small]{caption}
\usepackage{subcaption}
\usepackage{slashbox}
\usepackage{enumitem}

\hypersetup{breaklinks=true}

\makeatletter

\renewcommand{\p@enumi}{theenumi-}
\renewcommand{\@fnsymbol}[1]{\@arabic{#1}}

\newcommand{\fn}{\footnote}
\makeatother
\newtheorem{lemma}{Lemma}[section]

\newtheorem{example1}[lemma]{Example}
\newtheorem{ex1}[lemma]{Example}
\newtheorem{rem1}[lemma]{Remark}

\newtheorem{alg1}[lemma]{Algorithm}
\newtheorem{me1}[lemma]{Mechanism}

\usepackage{color}
\numberwithin{equation}{section}
\numberwithin{figure}{section}
\numberwithin{table}{section}

\usepackage[nomessages]{fp}
\usepackage[group-separator={,},group-minimum-digits=4]{siunitx}
 
\newcommand{\cost}[1]{\FPeval{\gbp}{round(336.05*#1,0)}\FPeval{\usd}{round(493*#1,0)}\num{#1} Galleons (\pounds\num{\gbp}, \$\num{\usd})} 

\begin{document}
\title{\vspace{-50pt}Harry Potter and the Goblin Bank of Gringotts\vspace{-10pt}}
\author{Zachary Feinstein\footnote{Zachary Feinstein, ESE, Washington University, St. Louis, MO 63130, USA, {\tt zfeinstein@ese.wustl.edu}.}\\ \textit{Washington University in St. Louis}}
\date{\today\fn{No funding, from either magical or muggle sources, was provided for this work.}}
\maketitle
\vspace{-20pt}
\addtocounter{footnote}{2}
\begin{abstract}
Gringotts Wizarding Bank is well known as the only financial institution in all of the Wizarding UK as documented in the works recounting the heroics of Harry Potter.  The concentration of power and wealth in this single bank needs to be weighed against the financial stability of the entire Wizarding economy.  This study will consider the impact to financial risk of breaking up Gringotts Wizarding Bank into five component pieces, along the lines of the Glass-Steagall Act in the United States.  The emphasis of this work is to calibrate and simulate a model of the banking and financial systems within Wizarding UK under varying stress test scenarios simulating rumors of Lord Voldemort's return or the release of magical creatures into an unsuspecting muggle populace.  We conclude by comparing the economic fallout from financial crises under the two systems: (i) Gringotts Wizarding Bank as a monopoly and (ii) the split-up financial system.  We do this comparison on the level of minimal system-wide capital injections that would be needed to prevent the financial crisis from surpassing the damage caused by Lord Voldemort.
\end{abstract}\vspace{-8pt}

\vspace{-18pt}
\section{Introduction}\label{Sec:Intro}
\vspace{-10pt}
On May 1, 1998, a trio of teenagers, with the aid of a former goblin employee, successfully broke into and robbed a vault at Gringotts Wizarding Bank \cite{HPbreakin}.  In this robbery, a priceless artifact -- Helga Hufflepuff's cup containing a piece of Lord Voldemort's soul -- was stolen and subsequently destroyed.  The current Minister of Magic, Hermione Granger, was one of those teenagers.\fn{Hermione Granger is elected Minister of Magic sometime before 2019 \cite{HPgranger}.  Before that time she is the heir apparent and top employee under Kingsley Shacklebolt.} 

Minister Granger thus knows of the flaws in having a single, monopolistic, financial institution.  Having a single bank, without competition, can hinder innovation.  In particular, without competition to drive innovation in, e.g.\ banking security, the once impenetrable vaults of Gringotts Wizarding Bank were breached twice in the 1990s.  Of course, more than the security concerns, the importance of a single institution concentrates too much (political) power in too few hands.  This could justify the reasoning that Wizarding UK's banking laws allow Bellatrix Lestrange to keep dangerous artifacts with impunity, and without testing, even after being given a life sentence at Azkaban.  Additionally, the mere appearance of the supreme importance of Gringotts Wizarding Bank has led the Ministry of Magic to nationalize Gringotts at various times, as recently as 1998 \cite{HPgringotts}.  As Minister Granger is a well-known proponent of the freedom of all magical creatures, the author of this study has been led to believe that the Minister is in favor of keeping Gringotts under goblin control.  To guarantee this freedom from Ministry control into the future may require reducing the scale of the bank.  Finally, with Hermione Granger's ties to the Muggle World both as a muggle-born witch and through conversations with the Prime Minister of Great Britain, she would have seen the dire consequences of financial crises and the proposed regulatory responses.

However, Hermione Granger was always one of the smartest witches or wizards of her generation.  She would not implement any new regulation without thorough study.  In particular, since Gringotts Wizarding Bank is a monopoly, Minister Granger has instructed the author of this paper to study the financial implications of breaking Gringotts into multiple pieces.\fn{Reportedly after unenlightening discussions with Professor Sybill Trelawney, who kept predicting death in every scenario.} Even further, Hermione Granger would remove Gringotts's power as the central bank and mint, returning such powers to the public.  Each of these pieces of Gringotts would be much weaker than the monopoly that existed beforehand, and thus the failure of one Baby Goblin\fn{Presuming the Wizarding World names broken up monopolies following the tradition of AT\&T.} would have less impact to the Wizarding World.  As a consequence there would be less chance of adverse incentives (such as moral hazard) from being too-big-to-fail.  But, there are potential negative consequences in such a scheme as well.

As the author of this study is a muggle\fn{Despite the protestations that ``Yer a wizard Harry'' from a rather large gentleman named Hagrid who seems very confused about the author's name.} and does not have access to magical documents that have the necessary economic data,\fn{A figure by the name of Arthur Weasley offered his aid, though after Weasley's third cycle of questions about the purpose of rubber ducks, the author had to cut off communication to finish the report on schedule.} we will first calibrate a model of the economy in Wizarding UK.  With this we will deduce the value of assets of Gringotts Wizarding Bank.  Then we can run stress testing and mathematical analysis on financial stability with either a single banking institution or multiple smaller institutions to determine a recommendation to the Ministry of Magic.

\vspace{-18pt}
\section{Modeling the Economics of the Wizarding World}\label{Sec:Model}
\vspace{-10pt}
It is well known, even amongst muggles, that the Wizarding economy is run on the Galleon (currency code WZG for Wizarding Galleons).  To analyze such an economy, we will want to compare economics to the muggle world.  In 2001, the official exchange rate offered by Gringotts Wizarding Bank was approximately 5.01 GBP/WZG, i.e.\ 5.01 British Pounds to 1 Wizarding Galleon, or 7.35 USD/WZG, i.e.\ 7.35 US Dollars to 1 Wizarding Galleon, using the exchange rate of 1.467 USD/GBP on March 12, 2001 (\cite{JKRowling-currency} and which is verified in \emph{Fantastic Beasts and Where to Find Them} and \emph{Quidditch Through the Ages} to give a value of 5.01 GBP/WZG (7.35 USD/WZG) \cite{wizard-money}).  As a direct result we can deduce that each Galleon contains at most 0.7648 grams of gold to prevent arbitrage opportunities (given a gold price of \$9.61 per gram on March 12, 2001) \cite{wizard-money}.  That is, wizards cannot obtain a risk-free profit by smelting the Galleon for gold, selling that for British pounds, and trading back into Galleons for more than was originally held.\fn{We assume that each Galleon contains a certain amount magic to prevent counterfeiting, thus the Galleon need not contain exactly 0.7648 grams of gold per coin.}

However, the official exchange rate of 5.01 GBP/WZG (or 7.35 USD/WZG on March 12, 2001) does not necessarily tell the whole economic story.  As a muggle in the UK cannot purchase any Wizarding goods, this imposed exchange rate would allow for squibs and muggle-born witches and wizards to function in the Wizarding World.  That is, the official exchange rate is utilized primarily as good-will via Gringotts and the Ministry of Magic.  Gringotts Wizarding Bank, a private institution, would be agreeable to this arrangement because it has the prominent public role as the central bank setting monetary policy.  This requires immense public trust and, as such, it must occasionally act in the public interest, and against its own interests, to justify the power it has been given.

To evaluate the true exchange rate via the notion of purchasing power parity, we consider the price of the \emph{Daily Prophet}, a Wizarding newspaper.  As reported in \cite{HPdailyprophet}, one newspaper could be bought for 5 Knuts at wholesale (in \emph{Harry Potter and the Sorcerer's Stone}) and for 1 Knut at a subscription price (evidenced in \emph{Harry Potter and the Order of the Phoenix}).  We will thus roughly estimate an exchange rate of \$1 per Knut \cite{HPdailyprophet}.  With 493 Knuts to the Galleon, this implies an exchange rate, utilizing a Newspaper Index,\fn{This is in lieu of the Big Mac Index \cite{BigMac}, as it appears witches and wizards do not partake in fast food.  We believe a newspaper to be an accurate index for purchasing power as magic would not sufficiently ease the investigatory costs of running a news organization, and as such the costs should be comparable.} of roughly 493 USD/WZG or 376.90 GBP/WZG in August 22, 2016 exchange rates \cite{HPwikibook}.

We do not care about the exchange rate between British Pounds and Wizarding Galleons for its own sake.  Instead we care so as to allow for accurate modeling of the Wizarding economy on the British Isles to both calibrate the size of Gringotts Wizarding Bank and consider the impacts to financial stability during a financial crisis if Gringotts were to be split up.  To deduce the size of the Wizarding economy we will assume that the percentage of Gross Domestic Product [GDP] used for education in Wizarding UK is comparable to that in (Muggle) UK.  As Hogwarts School for Witchcraft and Wizardry is the only government-sanctioned school in all of Wizarding UK \cite{PottermoreSchools},\fn{Wizarding UK may include the entirety of Ireland \cite{HPireland}.} and as the entirety of student tuition is covered by the Ministry of Magic \cite{JKtuition}, if we deduce the cost of attending Hogwarts we can find an estimate for the size of the economy in Wizarding UK.  

As described in \cite{HPhogwarts}, ``Hogwarts is considered to be one of the finest magical institutions in the Wizarding World.''  And as a boarding school, we can thus compare the tuition to other elite boarding schools in the UK.  For British students in the 2016-2017 school year, these elite boarding schools cost between \pounds35,000/year and \pounds40,000/year in the UK \cite{EtonCost,DulwichCost,QECost}.\fn{The cost at Eton is \pounds37,062/year~\cite{EtonCost}, Dulwich is \pounds39,480/year~\cite{DulwichCost}, Queen Ethelburga's (which the author has been assured is not magical, though the name seems to indicate otherwise) is approximately \pounds35,000/year~\cite{QECost}.}  Taking the midpoint, we will assume a total cost of roughly \pounds37,500/year for each student.  At the official exchange rate, the headline price of Hogwarts is 7,500 Galleons/year.  We utilize the official exchange rate since the Ministry of Magic pays the full cost of Hogwarts tuition for young witches and wizards \cite{JKtuition}.  As such, the price given would likely be with respect to the official exchange rate so as to keep up the pretense that a fair exchange rate has been provided.  This leads to a \emph{true} cost of \cost{7500} after adjusting for purchasing power via our Newspaper Index.  As Hogwarts ostensibly lasts 7 years, without any increases in tuition, this would cost \cost{52500} for a complete magical education.  While these are incredible tuition costs, recall that the tuition at Hogwarts is paid entirely by the Ministry of Magic \cite{JKtuition} and thus available to all qualified young witches and wizards.

As reported in \cite{BeyondHogwarts}, there are roughly 1,000 students per year who attend Hogwarts School for Witchcraft and Wizardry.  With our tuition costs from above, this leads to \cost{7500000} in revenue from tuition per year for Hogwarts.  And as Hogwarts is the sole government-sanctioned school in Wizarding UK \cite{PottermoreSchools} this cost is approximately going to be the entirety of the education budget for the Ministry of Magic.\fn{While Hogwarts is an elite boarding school, witches and wizards are, by definition, members of the elite.  Therefore all witches and wizards in the UK go to Hogwarts.}  Assuming similar profile to Muggle UK, the education budget is roughly 4.4\% of GDP \cite{UKEducation}, this leads to a Wizarding UK GDP of \cost{170454545}.  With approximately 10,000 wizards in all of the UK and Ireland \cite{Fictionomics}, this leads to an estimate of \cost{17045.45} GDP per capita for Wizarding UK.\fn{Normalized to GDP/capita and with the GDP/capita in the United States in 2015 is \$55,836.80, the 7,500 Galleon tuition at Hogwarts would be equivalent to \$24,568.19.  This is only \$5,000 more than in-state tuition (include room and board) at public universities within the United States \cite{USTuition}.}

Finally, we can deduce the wealth and size of Gringotts Wizarding Bank.  As the sole bank in all of Wizarding UK, as well as acting as the central bank for the Ministry of Magic, Gringotts would be of staggering size in relation to its economy.  To deduce this size, we will compare Gringotts Wizarding Bank to the size of the entire UK banking sector from a historical perspective.  In particular, before the rapid growth in the UK banking sector beginning in the 1970s, the financial sector in the United Kingdom had assets of approximately 100\% of of GDP \cite{qb14q402}.  Assuming the banking sector in Wizarding UK is of similar size, this leads us to conclude that Gringotts Wizarding Bank has assets of \cost{170454545}.\fn{Compare to JP Morgan Chase which has total assets of \$2.424 trillion \cite{WikiJPMorgan}, which while larger in absolute size is smaller as a fraction of the US GDP (approximately 13\%).}$^\text{,}$\fn{If the current size of the UK financial sector were used instead then Gringotts would have assets of approximately 450\% of GDP, i.e.\ \cost{767045452}.}

\vspace{-18pt}
\section{The Impact to Financial Stability from Splitting Up Gringotts}\label{Sec:RiskMsr}
\vspace{-10pt}
With the immense size of Gringotts with assets totaling 100\% of GDP, and with the vast scope of its operations, we will consider the implications of splitting it into 5 pieces.  Each so-called Baby Goblin bank will take a fraction of the assets and comprise some part of the banking sector.\fn{Motivated by the Glass-Steagall Act of 1933 in the United States separating commercial and investment banks.}  A quick summary of these five banks is as follows:
\begin{enumerate}[leftmargin=\parindent,noitemsep,topsep=0pt,parsep=0pt,partopsep=0pt]
\item \textbf{Bank of Gringotts [BofG]}: This institution retains the role as a \emph{central bank} for the Wizarding UK, and in such a position acts similarly to the Bank of England in the UK or the Federal Reserve Bank in the US.  The Bank of Gringotts retains the Gringotts name as the author feels this will give the most prestige to the founder Gringott.  A noninclusive list of all its responsibilities include: issuing banknotes, setting and maintaining the official exchange rate with Muggles, converting currency at that exchange rate, setting interest rate policy, and banking supervision.  Due to the high possibility of a conflict of interests, the BofG will not act as a depository or investment bank for witches, wizards, or their businesses.
\item \textbf{Keeper Wizarding Bank [KWB]}: This institution retains the roles of a \emph{commercial and retail bank}.  The Keeper Wizarding Bank would act as the depository institution for most witches, wizards, and small businesses.  It will run the current Gringotts vaults as well as give the savings accounts and loans (e.g., mortgages and small business loans) that are traditionally handled by banks and credit unions.  The KWB would provide simple banking in a low-risk and low-reward environment.
\item \textbf{Seeker Wizarding Bank [SWB]}: This institution retains the role of an \emph{investment bank}.   The Seeker Wizarding Bank would act as an issuer of financial securities, raising of capital on behalf of corporations, and facilitating mergers and acquisitions.  The SWB will additionally act as a market maker and trade various securities itself.  The SWB will keep larger corporations fully capitalized, and provide higher returns for investors (though with the requisite larger risks to compensate).
\item \textbf{Chaser Wizarding Fund [CWF]}: This institution retains the role of a \emph{hedge fund}.  The Chaser Wizarding Fund would pool investments made by witches and wizards into a complex portfolio.  It will be highly levered (i.e., the CWF will invest with borrowed Galleons) and thus extremely risky, but in exchange can provide very high returns.  The positions the CWF takes can be long (purchasing assets) or short (selling assets) in order to hedge market risk. 
\item \textbf{Beater Wizarding Insurance Group [BWIG]}: This institution retains the role of an \emph{insurance company}.  The Beater Wizarding Insurance Group would provide the typical range of insurance options for witches and wizards.  This includes property insurance (for example homeowners insurance), life insurance, and transportation insurance (be it via apparition, Floo network, portkey, broom, or enchanted flying automobile).  The BWIG also provides insurance for the other Baby Goblins, for instance default insurance. 
\end{enumerate}

\captionsetup[wrapfigure]{skip=3pt}
\begin{wrapfigure}{R}{0.43\textwidth}
\centering
\psset{unit=0.6cm}
\begin{pspicture}(-6,-4)(6,4)
\psellipse(0,0)(1,0.4)
\rput(0,0){BWIG}
\psellipse(-5,0)(1,0.4)
\rput(-5,0){CWF}
\psellipse(5,0)(1,0.4)
\rput(5,0){BofG}
\psellipse(5,3)(1,0.4)
\rput(5,3){KWB}
\psellipse(5,-3)(1,0.4)
\rput(5,-3){SWB}
\psline[linewidth=.6mm]{<->}(-4,0)(-1,0)
\psline[linewidth=.6mm]{<->}(1,0)(4,0)
\psline[linewidth=.6mm]{<->}(.8,.2)(4.2,2.8)
\psline[linewidth=.6mm]{<->}(.8,-.2)(4.2,-2.8)
\psline[linewidth=.6mm]{<->}(5,0.4)(5,2.6)
\psline[linewidth=.6mm]{<->}(5,-0.4)(5,-2.6)
\psline[linewidth=.6mm]{<->}(-4.2,.2)(4,3)
\end{pspicture}
\caption{Network of interbank liabilities between the Baby Goblins.  Does not include the Ministry of Magic or Wizarding society at large.}
\label{Fig:Network}
\end{wrapfigure}
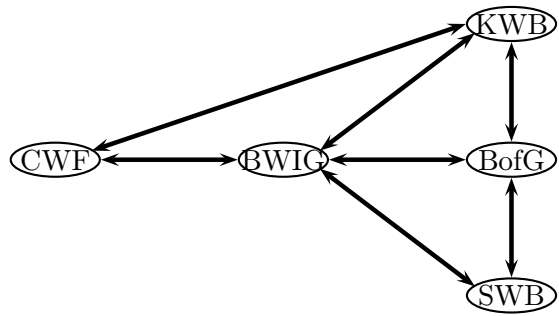
In describing the financial system we will additionally consider the 10,000 witches and wizards in Wizarding UK as well as the Ministry of Magic itself.  These will be added to our system as an aggregated node, which we will call \textbf{society} for lack of a better term.  The societal node will be connected to all the Baby Goblins (except the Bank of Gringotts) as citizens and the Ministry of Magic will have financial liabilities or deposits with each institution. 
The connections between the Baby Goblins are displayed in Figure~\ref{Fig:Network}.  
By dividing Gringotts Wizarding Bank 
into the five component institutions previous intrabank obligations become interbank liabilities.  In particular, due to the separation of the institutions, the Baby Goblins will not implicitly bail each other out in times of failures.  However, having the map of interbank liabilities is insufficient to fully describe our system.  As we have split up Gringotts Wizarding Bank into 5 pieces based on their function, they need not have equal size.

The Bank of Gringotts, being the central bank, will likely have assets similar in size to the Bank of England on average.  Historically from 1696 to 2014, the Bank of England has held 11\% of nominal GDP in total assets \cite{BofEbalance}.  Thus we will assume that the BofG will hold 11\% of Wizarding GDP as assets.  As a simplification, we will assume that the other four institutions split the remaining 89\% of assets equally amongst themselves.  The size of nominal interbank liabilities is displayed in Table~\ref{Table:Network}.  As we split the assets equally amongst the 4 private banks, we will similarly split the total 60\% of nominal GDP in obligations to society.\fn{As a significant amount of assets are in the vaults the fraction of assets to liabilities is assumed to be small.}$^\text{,}$\fn{In the case that Gringotts Wizarding Bank is not split into the component pieces it has 100\% of nominal GDP in total assets and 60\% of nominal GDP in liabilities to society.}

\begin{table}
\centering
\begin{tabular}{|r||c|ccccc|}
\hline
\backslashbox{From}{To} & Society & BofG & KWB & SWB & CWF & BWIG \\\hline\hline
Assets & * & 11 & 22.25 & 22.25 & 22.25 & 22.25 \\\hline
BofG & * & * & 7.5 & 5 & * & 7.5 \\
KWB & 15 & 2 & * & * & 3 & 7.5 \\
SWB & 15 & 2 & * & * & * & 5 \\
CWF & 15 & * & 5 & * & * & 5 \\
BWIG & 15 & 2 & 5 & 5 & 5 & * \\\hline
\end{tabular}
\caption{Summary of assets and liabilities; the numbers denote the amount that the ``row institution'' owes to the ``column institution'' as a \% of GDP.}
\label{Table:Network}
\end{table}

\begin{wrapfigure}{R}{0.60\textwidth}
\centering
\includegraphics[width=0.59\textwidth]{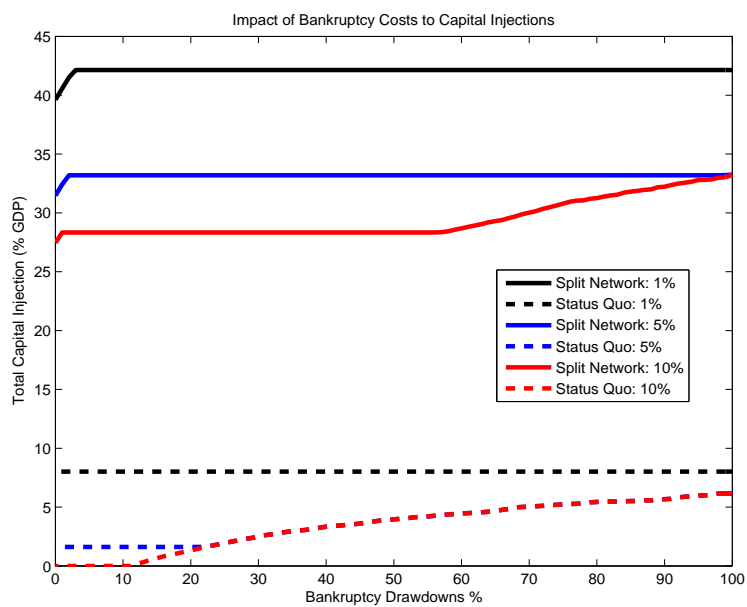}
\caption{The impact of bankruptcy costs on total capital injections imposed on the system.  Solid lines indicate the 5-bank system; dashed lines indicate the status quo system with the singular Gringotts Wizarding Bank.}
\label{Fig:Drawdown}
\end{wrapfigure}
We now wish to compare these two financial systems (the status quo and the 5-bank system) under varying stress scenarios.  For our purposes, we will consider the model of \cite{RV11} which includes drawdowns conditional on bankruptcy (i.e.\ bankruptcy costs) caused by, e.g., rumors of the return of Lord Voldemort or muggles discovering Wizarding society.  For comparison purposes we will consider the minimal amount of Galleons to be provided to the banks by the Ministry of Magic ``today'' in order to compensate for the risk of a crisis ``tomorrow'' (henceforth called the \emph{minimal cost of capital injections}) defined via the efficient allocation rules in \cite{feinstein2014measures}.  By considering the minimal cost of capital injections rather than the full risk measures we can more easily compare the results of our stress testing of the full system.\fn{Of course allowing for a large enough supply of Felix Felicis, the entire notion of a financial crisis could be put off indefinitely.  Though if the supply of that powerful potion were ever interrupted, this would cause a dramatic correction in the market.}  
To aggregate the impact to society, we consider losses caused by defaulting banks; this is caused by both unpaid liabilities to society and negative equity to the Bank of Gringotts (in the split-up case) or Gringotts Wizarding Bank (under the status quo scenario).  We consider three criteria that may be used to determine the size and composition of the necessary capital injections: so that the expected losses to the Wizarding economy in the worst 1\% (5\% or 10\% respectively) of cases are no worse than 1\% of GDP (i.e., \cost{1704545.45}).  

\begin{wrapfigure}{L}{0.60\textwidth}
\centering
\includegraphics[width=0.59\textwidth]{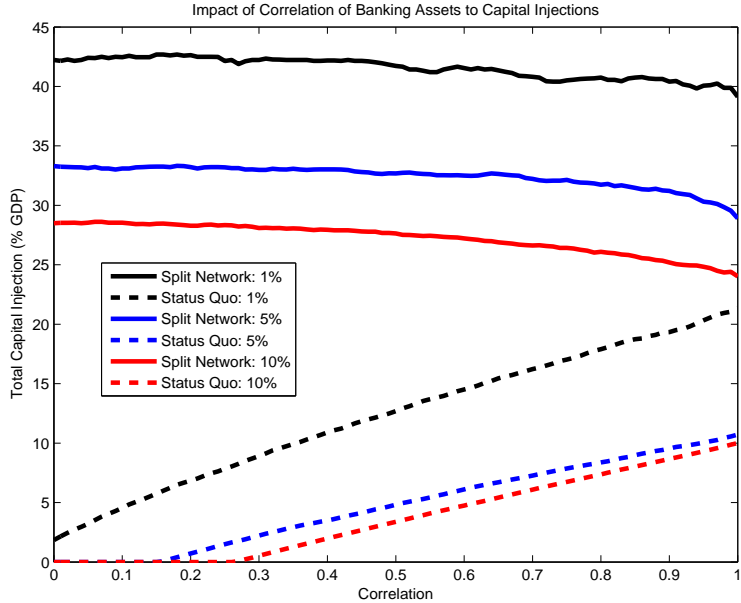}
\caption{The impact of asset correlations on total capital injections given to the system.  Solid lines indicate the 5-bank system; dashed lines indicate the status quo system with the singular Gringotts Wizarding Bank.}
\label{Fig:Corr}
\end{wrapfigure}
\begin{wrapfigure}{R}{0.60\textwidth}
\centering
\includegraphics[width=0.59\textwidth]{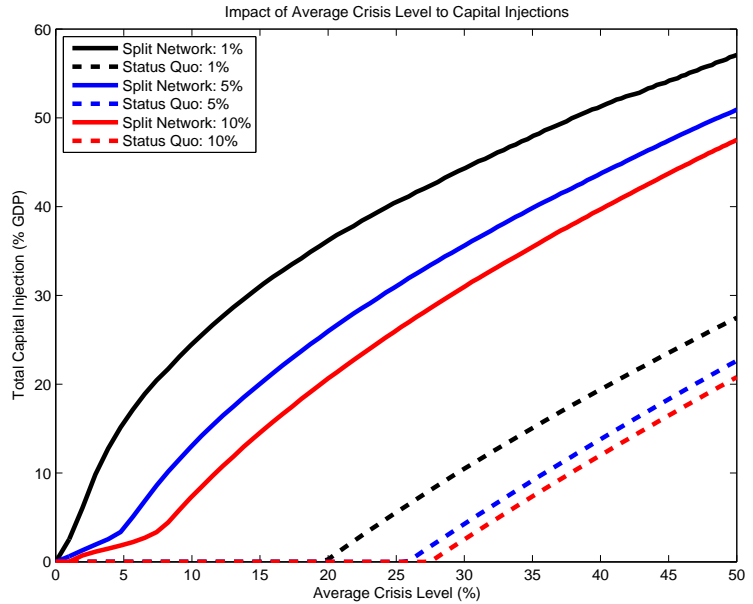}
\caption{The impact of stress scenarios on total capital injections given to the system.  Solid lines indicate the 5-bank system; dashed lines indicate the status quo system with the singular Gringotts Wizarding Bank.}
\label{Fig:Dist}
\end{wrapfigure}

To consider stressing the system, we analyze possible crises that would result from situations such as a threat of muggles discovering information on the Wizarding World (e.g.\ through the accidental release of magical creatures) or rumors of Lord Voldemort's return (e.g.\ if a new time turner were to be found).  Such an event would panic the market and causing a rush to safety as individuals and businesses alike attempt to withdraw hard assets such as golden Galleons.  This would precipitate a firesale and thus a drop in banking asset values.\fn{All stresses are mathematically defined via beta distributed random variables multiplied by the asset values; correlations are included through the use of Gaussian copulas.}  As we can split the assets amongst the Baby Goblins in different ways,\fn{To deduce the wealth of Gringotts Wizarding Bank, we assume the same stresses as the Baby Goblins and then take the summation to recreate the merged institution.} we will consider the case in which bank asset correlations may vary.\fn{The results are displayed in Figure~\ref{Fig:Corr} under 10\% bankruptcy costs and a 27\% (average) drop in asset values.}  Similarly, though we know that muggles finding out about the Wizarding World or rumors of Lord Voldemort's return would cause a financial crisis, the strength of those threats could vary significantly; as such we will also consider the case in which the shock sizes are varied.\fn{The results are displayed in Figure~\ref{Fig:Dist} under 10\% bankruptcy costs and a 25\% correlation of assets.}

As expected, as drawdowns or stress increases so does system stability in each of the three rules we consider under both the status quo and split up scenarios (see Figures~\ref{Fig:Drawdown} and~\ref{Fig:Dist} respectively).  By varying the bankruptcy costs only, the capital injections for the status quo monopolistic case with only Gringotts Wizarding Bank requires between approximately \cost{41000000} and \cost{58000000} less than the split up case with the Baby Goblins.  While changing the crisis level, in the most extreme case the status quo requires approximately \cost{61000000} less in capital injections than the Baby Goblins would.  Interestingly, increasing asset correlation, while decreasing stability for Gringotts Wizarding Bank appears to increase stability for the system of Baby Goblins (see Figure~\ref{Fig:Corr}).  Though here too the status quo saves between approximately \cost{24000000} and \cost{69000000} in capital injections over the Baby Goblins.   What is very noticeable in each of the considered scenarios is that the status quo, in which Gringotts Wizarding Bank is kept whole, produces lower risk levels than the split up case with 5 Baby Goblins.

In fact, this can be shown to be the case in any situation we wish to consider.  It is proven in \cite{RV11} that, with bankruptcy costs, there is an incentive for mergers if there are failures in the system.\fn{This was seen during the 2008 financial crisis when Bear Stearns was bought by JP Morgan Chase or Merrill Lynch by Bank of America.}  That is, during a financial crisis, after splitting up Gringotts Wizarding Bank, the individual institutions (BofG, KWB, SWB, CWF, and BWIG) would wish to merge back together to maximize individual profits and reduce aggregate risk.  This gives us a definite result: splitting up Gringotts Wizarding Bank would increase the potential for systemic crises.

\vspace{-18pt}
\section{Conclusion}\label{Sec:Conclusion}
\vspace{-10pt}
We calibrated a model of the Wizarding economy in late 2016 in order to determine the risks posed by the too-big-to-fail Gringotts Wizarding Bank.  This allowed us to consider the counterfactual scenario in which this systemically important financial institution has been broken up into five component pieces, each taking a piece of the financial business.  Under varying stress scenarios, we considered the impact to the Wizarding economy that splitting up Gringotts Wizarding Bank would have.  This allows us to determine the comparative impact that breaking up the bank would have on economic and financial stability.  The conclusion of this author is that, under the model presented in \cite{RV11}, the status quo promotes economic wellbeing and reduces financial risk.  However, there may be other political and economic reasons, such as promoting innovation and reducing adverse incentives (e.g.\ moral hazard), which might overwhelm the financial stability results presented herein.

{\footnotesize
\bibliographystyle{plain}
\bibliography{bibtex2}
}

\end{document}